\documentclass[10pt]{article}

\usepackage{amsmath}
\usepackage{amsfonts}
\usepackage{amssymb}
\usepackage{amsthm}
\usepackage{stmaryrd} 
\usepackage{braket}
\usepackage{empheq}
\usepackage{graphicx}
\usepackage{subfigure}
\usepackage{hyperref}
\usepackage{footmisc}
\usepackage{appendix}
\usepackage{epsf}
\usepackage{epsfig}
\usepackage{tikz,tikz-3dplot}
\usetikzlibrary{decorations.markings,decorations.pathmorphing}
\usepackage{caption}
\usepackage{pgfplots}
\usepackage[utf8]{inputenc}
\usepackage[english]{babel}
\usepackage[T1]{fontenc}
\usepackage{sectsty}
\usepackage{authblk}
\subsubsectionfont{\normalfont\itshape}

\usepackage{fullpage}

\renewcommand{\leq}{\leqslant}

\newcommand{\bbone}{{\text{\usefont{U}{bbold}{m}{n}\char49}}}

\theoremstyle{plain}
\newtheorem{theorem}{Theorem}[section]
\newtheorem{proposition}[theorem]{Proposition}

\theoremstyle{definition}
\newtheorem{definition}[theorem]{Definition}

\title{Disentangling tensor product structures}
\date{}

\author[1 2]{\textsc{Antoine Soulas}}
\affil[1]{Quantum Optics, Quantum Nanophysics and Quantum Information, Faculty of Physics, University of Vienna, Vienna, Austria \vspace{0.2cm}} 
\affil[2]{Institute for Quantum Optics and Quantum Information (IQOQI), Austrian Academy of Sciences, Vienna, Austria \vspace{0.2cm}}

\begin{document}
\maketitle

\vspace{-1.5cm}

\begin{center}
  \rule{6cm}{1pt}
\end{center}

\abstract{As a contribution to the field of quantum mereology, we study how a change of tensor product structure in a finite-dimensional Hilbert space affects its entanglement properties. In particular, we ask whether, given a time-evolving state, there exists a tensor product structure in which no entanglement is generated. We give a concrete, constructive example of disentangling tensor product structure in the case of a C-NOT gate evolution between two qbits, before showing that this cannot be achieved for most time-evolving quantum states.}


\vspace{0.4cm}
\hrule

\tableofcontents


\section{Introduction}
Although not new, the field of \textit{quantum mereology} has attracted some attention lately. In a nutshell, it is concerned with the decomposition of any quantum system into subsystems and with the study of the tensor product structure (TPS) of its Hilbert space. Since the notion of locality and no-signalling properties are so deeply rooted in the TPS \cite{kennedy1995empirical}, its study might have very far-reaching consequences. In particular, several authors have suggested how spacetime itself could in turn emerge from the TPS \cite{cao2017space, carroll2019mad, carroll2022reality, giddings2019quantum, piazza2010glimmers}. It also have links with the quantum-to-classical transition: as stated in \cite{loizeau2025quantum}, ‘the emergence of classicality may be a tensor structure problem’, an idea also put forward in \cite{carroll2021quantum, adil2024search, franzmann2024or}. 

\textit{A priori}, there exists an infinity of mathematically admissible TPSs, most of which clearly do not correspond to the one we routinely use in quantum physics, nor to the decomposition of the world we perceive. So which one is ours? \cite{stoica2023prince, stoica2025makes} A general aim in quantum mereology is to be able to select intrinsically or dynamically a preferred TPS given a Hamiltonian plus possible additional requirements. 

A variety of ideas have been proposed to single out a TPS, in other words to characterize the decomposition into subsystems that we actually perceive. These include: decoherence-induced decompositions and approximate pointer states \cite{carroll2021quantum, adil2024search}, locality of the Hamiltonian (although the result of the main reference supporting this approach \cite{cotler2019locality} has been widely misinterpreted as a result on the strict uniqueness of the TPS given a Hamiltonian plus some locality condition, but such a claim can not hold \cite{stoica20213d, stoica2023no, stoica2024does}; in fact, \cite{cotler2019locality} shows unicity up to a global unitary equivalence relation; see \cite{soulas2025on} for a thorough investigation of this topic), analytic properties of the Hamiltonian's spectrum \cite{loizeau2025quantum}, minimal scambling \cite{zanardi2024operational}, maximal integration \cite{tegmark2015consciousness}, single measurement outcomes \cite{franzmann2024or}. 

Other important recent works include the observation that typical Hamiltonians can generally be well-approximated by $2$-local Hamiltonians in some TPS \cite{loizeau2023unveiling}, and the geometric study of the behaviour of a TPS under unitary evolution presented in \cite{andreadakis2025tensor}. The latter is based on Zanardi \textit{et al.}'s theorem \cite{zanardi2004quantum}, stating that a choice of TPS is isomorphic to a choice of commuting subalgebras generating $\mathcal{L}(\mathcal{H})$. Finally, the notion of subsystem locality has also been investigated from a quantum reference frame (QRF) framework, hinting at a frame-dependent notion of subsystems \cite{ali2022quantum}. 

Our aim is to contribute to this growing literature with a new perspective. It is well-known that one can always find a TPS such that the entanglement properties of any pure state can be tailored \cite{liu2004relative, harshman2011observables}. In addition, for a general time-evolving quantum state, one can trivially construct a time-evolving TPS that ‘follows’ the state so that no entanglement ever occur \cite{schwindt2012nothing}. In this paper, we propose to go further and ask the following question: given a time-evolving quantum state, can we find a \textit{fixed} TPS that ‘disentangles’ the whole evolution? This problem is partly motivated by some considerations of \cite[\S8.2]{franzmann2024or}, namely that entanglement might bear some similarities with gravity, \textit{i.e.} a phenomenon that appear when choosing a ‘wrong’ TPS, respectively a ‘wrong’ (non-inertial) set of coordinates.

We start with some definitions and a proper statement of the problem in Sec. \ref{definitions}. An explicit constructive example of a disentangling TPS is given in Sec. \ref{existence} in the case of a C-NOT gate evolution between two qbits. Next, we show in Sec. \ref{non-existence} that what could be achieved in the C-NOT gate example is impossible in most cases.  Finally, we explore in Sec. \ref{hamiltonian} the same problem from the point of view of the Hamiltonian.

\section{Definitions} \label{definitions}
In the sequel, $\mathcal{H}$ always denotes a complex Hilbert space of dimension $n = n_1 n_2$. Note that, in this work, we will restrict ourselves to finite dimensional spaces, pure states and bipartite tensors products. 

\begin{definition}[Tensor product structure]
A (bipartite) \textit{TPS} of $\mathcal{H}$ is an equivalence class of isomorphisms $\Phi : \mathcal{H} \rightarrow \mathcal{H}_1 \otimes \mathcal{H}_2$ that factorize $\mathcal{H}$ into 2 blocks of respective dimension $n_1$ and $n_2$, where two isomorphisms $\Phi_1$ and $\Phi_2$ are said to be equivalent if $\Phi_1 \Phi_2^{-1}$ is a product of local unitaries $U_1 \otimes U_2$.
\end{definition}

In particular, picking a basis $\mathcal{B}$ of $\mathcal{H}$, and labelling its elements by two indices ranging from 1 to $n_1$ and 1 to $n_2$ respectively, as $\mathcal{B} = ( \ket{11} , \dots, \ket{n_1 n_2} )$, obviously provides an isomorphism $\Phi$, and therefore unambiguously defines a TPS $[\Phi]$ of $\mathcal{H}$.

\begin{definition}[Disentangling tensor product structure]
Denote $\mathbb{S}^n$ the complex unit $n$-sphere (of real dimension $2n-1$) of $\mathcal{H}$. We call \textit{trajectory} in $\mathcal{H}$ the evolution of a normalized quantum state on $\mathbb{S}^n$, that is a continuous map:
\[ \begin{array}{lcll} & [0,T] &  \longrightarrow &  \mathbb{S}^n \\ & t & \longmapsto  & \ket{\Psi(t)}.  \end{array}\]
We say that a TPS $\mathcal{H} = \mathcal{H}_1 \otimes \mathcal{H}_2$ \textit{disentangles} a trajectory $\ket{\Psi(t)}$ if the latter is a product state at all times in this TPS, \textit{i.e.} if there exist two trajectories $\ket{\Psi_1(t)}$ and $\ket{\Psi_2(t)}$ in $\mathcal{H}_1$ and $\mathcal{H}_2$ such that $\ket{\Psi(t)} = \ket{\Psi_1(t)} \otimes  \ket{\Psi_2(t)}$ for all $t \in [0,T]$.
\end{definition}

Now, the mathematical question we want to consider is: can we find a TPS that disentangles a given trajectory in $\mathcal{H}$?

\section{Existence for the C-NOT gate} \label{existence}

The answer is obvious if ones allows the TPS to depend on time, as explained in \cite{schwindt2012nothing}. At each $t$, define $\ket{11(t)} = \ket{\Psi(t)}$ and complete it into a basis $( \ket{11(t)} , \ket{12(t)}, \dots, \ket{n_1 n_2 (t)} )$. In this TPS, $\ket{\Psi(t)} = \ket{1(t)} \otimes \ket{1(t)}$ is a non-entangled state for all $t$. \\

If now the TPS has to be fixed, things are more complicated. One can first notice that if
\[ \dim\Big( \mathrm{vect}\{ \ket{\Psi(t)} \mid t\in [0,T] \} \Big) \leq \max(n_1, n_2) \]
(say here $\max(n_1, n_2)=n_1$), \textit{i.e.} whenever the trajectory is contained in a sufficiently low dimensional subspace, we can take $(\ket{11}, \dots , \ket{ n_1 1})$ a basis of such a subspace, complete it into $( \ket{11} , \dots, \ket{n_1 n_2} )$, and we have indeed $\ket{\Psi(t)} = \ket{\Psi_1(t)} \otimes \ket{1}$ for all $t$. \\

Clearly, the condition above is not a necessary one, because it includes only very special trajectories which are constant on one side of the TPS, namely of the form $\ket{\Psi_1(t)} \otimes \ket{1}$. In general, given a TPS, how does the set of its product states $\{ \ket{\Psi_1} \otimes \ket{\Psi_2} \mid \ket{\Psi_1} \in \mathbb{S}^{n_1},  \ket{\Psi_2} \in  \mathbb{S}^{n_2} \} \subset \mathbb{S}^n$ looks like? It is a submanifold of $\mathbb{S}^n$ of real dimension $2(n_1+n_2)-3$, so it seems that there is quite some room, and maybe we can find a TPS for which the trajectory $\ket{\Psi(t)}$ always lies in this set. We will now prove that this is indeed possible in the simplest example of a C-NOT gate applied between two qbits. \\

\begin{proposition} \label{C-NOT}
Let $\mathcal{H} = \mathbb{C}^2 \otimes \mathbb{C}^2$ be the standard TPS of local observables associated with two qbits, and $\ket{\Psi(t)} = \frac{1}{\sqrt{2}} \begin{pmatrix}  1 \\ 0 \\ \cos(t) \\  \sin(t) \end{pmatrix}$  for $t \in [0, \frac{\pi}{2}]$ be the trajectory, as written in the basis $\mathcal{B} = (\ket{00}, \ket{01}, \ket{10}, \ket{11})$, describing a C-NOT gate evolution $\frac{1}{\sqrt{2}} (\ket{0} + \ket{1}) \otimes \ket{0} \rightsquigarrow \frac{1}{\sqrt{2}} (\ket{00} + \ket{11})$. There exist a TPS of $\mathcal{H}$ in which $\ket{\Psi(t)}$ is a product state for all $t \in [0, \frac{\pi}{2}]$.
\end{proposition}

\begin{proof}

\textit{1. Short answer.} Let's define a new TPS, obtained from the standard one by acting with the unitary 
\[U = \frac{1}{\sqrt{2}}  \begin{pmatrix} 
-1 & 0 & 1 & 0  \\ 
 0 & i & 0 & i \\ 
 0 & -i & 0 & i \\
 1 & 0 & 1 & 0 \end{pmatrix}, \]
said differently $U$ maps $\mathcal{B}$ to the basis $\mathcal{B}' = (\ket{++}, \ket{+-}, \ket{-+}, \ket{--})$ defining this new TPS. Transcripted in $\mathcal{B}'$, the trajectory $\ket{\Psi(t)}$ writes:
\begin{align*}
\frac{1}{\sqrt{2}} U \begin{pmatrix}  1 \\ 0 \\ \cos(t) \\  \sin(t) \end{pmatrix} 
&=\frac{1}{2} \begin{pmatrix}  -1+ \cos(t) \\ i \sin(t) \\ i \sin(t) \\  1+ \cos(t) \end{pmatrix} \\
&= \frac{1}{4} \begin{pmatrix}  -2+e^{it}+e^{-it} \\ e^{it}-e^{-it} \\ e^{it} - e^{-it} \\  2+e^{it}-e^{-it} \end{pmatrix} \\
&= \frac{e^{-it}}{4} \begin{pmatrix}  (e^{it}-1)^2 \\ (e^{it}-1)(e^{it}+1) \\ (e^{it} +1)(e^{it}-1) \\  (e^{it}+1)^2 \end{pmatrix} \\
&= \frac{e^{-it}}{4} \left[ (e^{it}-1) \ket{+} + (e^{it}+1) \ket{-} \right] \otimes \left[ (e^{it}-1) \ket{+} + (e^{it}+1) \ket{-} \right]. \end{align*}

Note, of course, that one can compose $U$ with any unitary of the form $V_1 \otimes V_2$ and preserve this property (but this is actually the same TPS).

\textit{2. Detailed construction.} Here is a way to find the above unitary. Assuming that such a TPS exists, let's first look for a necessary condition that constrains it. As before, denote $U = (u_{ij})_{1 \leq i,j \leq 4 }$ the unitary that maps $\mathcal{B}$ to the basis $\mathcal{B}' = (\ket{++}, \ket{+-}, \ket{-+}, \ket{--})$ of this new TPS. Note that upon replacing $\cos(t)=\frac{e^{it} + e^{-it}}{2}$ and $\sin(t)=\frac{e^{it} - e^{-it}}{2i}$, any expression of the form $\alpha + \beta \cos(t) + \gamma \sin(t)$ with $\alpha, \beta, \gamma \in \mathbb{C}$ can be rewritten as a trigonometric polynomial $e^{-it} P(e^{it})$, with $P \in \mathbb{C}[X]$ a polynomial of order 2. Transcripted in $\mathcal{B}'$, the trajectory writes:

\begin{equation} \frac{1}{\sqrt{2}} U \begin{pmatrix}  1 \\ 0 \\ \cos(t) \\  \sin(t) \end{pmatrix} =\frac{1}{\sqrt{2}} \begin{pmatrix}  u_{11} + u_{13} \cos(t) + u_{14} \sin(t) \\ u_{21} + u_{23} \cos(t) + u_{24} \sin(t) \\ u_{31} + u_{33} \cos(t) + u_{34} \sin(t) \\  u_{41} + u_{43} \cos(t) + u_{44} \sin(t) \end{pmatrix} = \frac{e^{-it}}{\sqrt{2}} \begin{pmatrix}  P_1(e^{it}) \\ P_2(e^{it}) \\ P_3(e^{it}) \\  P_4(e^{it}) \end{pmatrix}, \label{polynomials} \end{equation}

with $\deg(P_1) = \deg(P_2) = \deg(P_3) = \deg(P_4) = 2$.

Now, any product state in the new TPS takes the form $(a\ket{+} + b\ket{-}) \otimes(c\ket{+} + d\ket{-}) = \begin{pmatrix}  ac \\ ad \\ bc \\ bd \end{pmatrix}$ when written in $\mathcal{B}'$. Since $\ket{\Psi(t)}$ is by assumption a product state in this TPS, this implies in particular:
\begin{equation} \forall t \in [0, \frac{\pi}{2}], \quad P_1(e^{it}) P_3(e^{it}) = P_2(e^{it}) P_4(e^{it}). \label{product_state} \end{equation}
Decomposing into irreducible factors, there exists four degree 1 polynomials $A = X - \lambda_a$, $B = X - \lambda_b$, $C = X - \lambda_c$, $D = X - \lambda_d$ such that $P_1(e^{it}) P_3(e^{it}) = P_2(e^{it}) P_4(e^{it}) = \kappa A(e^{it})B(e^{it})C(e^{it})D(e^{it})$, where $\kappa = \kappa_1 \kappa_3 = \kappa_2 \kappa_4$ is the product of the leading coefficients of $P_1$ and $P_3$, resp. $P_2$ and $P_4$. In particular, each $P_i$ is, up to a constant factor, the product of two polynomials among $\{ A, B, C, D \}$.There are now two cases to distinguish. 

Either the roots of $P_1$ are exactly those of $P_3$ (resp. $P_4$) and the roots of $P_2$ are exactly those of $P_4$ (resp. $P_3$). In view of $\eqref{polynomials}$, this entails that $u_{11} + u_{13} \cos(t) + u_{14} \sin(t) = \eta ( u_{31} + u_{33} \cos(t) + u_{34} \sin(t))$ for some $\eta \in \mathbb{C}$, hence $(u_{11} , u_{13} , u_{14}) = \eta (u_{31} , u_{33} , u_{34})$, and similarly $(u_{21} , u_{23} , u_{24}) = \gamma (u_{41} , u_{43} , u_{44})$. But this is impossible, since such a $U$ cannot be an invertible matrix (develop its determinant with respect to the second row, and find a sum of vanishing $3\times3$ determinants, because there are always two collinear lines).

The second possibility is that the roots are intertwined, for instance (the other cases are treated similarly):
 \[ \left\{ \begin{array}{lll}
P_1 &= \kappa_1 A C &= \kappa_1 (X^2 - ( \lambda_a +  \lambda_c) X +  \lambda_a \lambda_c) \\
P_2 &= \kappa_2 A D &= \kappa_2 (X^2 - ( \lambda_a +  \lambda_d) X +  \lambda_a \lambda_d) \\
P_3 &= \kappa_3 BC &= \kappa_3 (X^2 - ( \lambda_b +  \lambda_c) X +  \lambda_b \lambda_c)  \\
P_4 &= \kappa_4 BD &=\kappa_4 (X^2 - ( \lambda_b +  \lambda_d) X +  \lambda_b \lambda_d).
\end{array} \right.  \]
Recall, on the other hand, that the $P_i$ were built from the $(u_{ij})_{1 \leq i,j \leq 4}$ as:
\[ P_i = \frac{u_{i3} - i u_{i4}}{2} X^2 + u_{i1} X + \frac{u_{i3} + i u_{i4}}{2}. \]
Identifying the coefficients, and inverting the equations, one finds that $U$must be of the form:
\[U = \begin{pmatrix}  -\kappa_1 (\lambda_a + \lambda_c)  & * & \kappa_1 (1 + \lambda_a \lambda_c) & i \kappa_1 (1 - \lambda_a \lambda_c)   \\ 
 -\kappa_2 (\lambda_a + \lambda_d)  & * & \kappa_2 (1 + \lambda_a \lambda_d) & i \kappa_2 (1 - \lambda_a \lambda_d) \\ 
 -\kappa_3 (\lambda_b + \lambda_c)  & * & \kappa_3 (1 + \lambda_b \lambda_c) & i \kappa_3 (1 - \lambda_b \lambda_c) \\
 -\kappa_4 (\lambda_b + \lambda_d)  & * & \kappa_4 (1 + \lambda_b \lambda_d) & i \kappa_4 (1 - \lambda_b \lambda_d) \end{pmatrix}. \]
 
We are therefore looking for some complex numbers $\kappa_1, \kappa_2, \kappa_3, \kappa_4, \lambda_a,  \lambda_b, \lambda_c,  \lambda_d$ such that $\kappa_1 \kappa_2 = \kappa_3 \kappa_4$ and the matrix above is unitary. Taking appropriate linear combinations, this boils down to finding three orthonormal vectors of the form:

\[\begin{pmatrix}  2\kappa_1 \\ 2\kappa_2 \\ 2\kappa_3 \\  2\kappa_4 \end{pmatrix} , \begin{pmatrix}  2\kappa_1 \lambda_a \lambda_c \\ 2\kappa_2 \lambda_a \lambda_d \\ 2\kappa_3\lambda_b \lambda_c \\  2\kappa_4 \lambda_b \lambda_d \end{pmatrix} ,  \begin{pmatrix}  2\kappa_1 (\lambda_a +\lambda_c) \\ 2\kappa_2 (\lambda_a +\lambda_d) \\ 2\kappa_3(\lambda_b +\lambda_c) \\  2\kappa_4 (\lambda_b+ \lambda_d) \end{pmatrix}. \]
It turns out that setting $\kappa_1 = \kappa_2 = \kappa_3 = \kappa_4 = \frac{1}{2\sqrt{2}}$, $\lambda_a=\lambda_c=1$ and $\lambda_b=\lambda_d=-1$ works and yields the unitary given in the short answer.
\end{proof}

\section{Non-existence in most cases} \label{non-existence}

Clearly, the above proof is very specific to the particular case of the C-NOT gate involving trigonometric functions, and can't be generalized. In fact, we will now prove that ‘most’ trajectories do not admit a disentangling TPS. \\

\begin{proposition} \label{general}
Let $\mathcal{H} = \mathbb{C}^n \otimes \mathbb{C}^m$ be a TPS of  $\mathcal{H}$ defined by a basis $\mathcal{B} = (\ket{1 1},\dots, \ket{n m})$. For $t \in [0, T]$ and some continuous complex functions $(a_{i,j}(t))_{\substack{1\leq i \leq n \\ 1 \leq j \leq m}}$, let $\ket{\Psi(t)} = \begin{pmatrix}  a_{1,1}(t) \\ a_{1,2}(t) \\ \vdots \\  a_{n,m}(t) \end{pmatrix}$ be an arbitrary trajectory on the sphere $\mathbb{S}^n$, as written in $\mathcal{B}$. If the $(a_{i,j} a_{k,l})$ are linearly independent in $\mathcal{C}^0([0,T])$, then there does not exist a TPS of $\mathcal{H}$ in which $\ket{\Psi(t)}$ is a product state for all $t$.
\end{proposition}

The hypothesis of the proof seems a very weak one. Because $\mathcal{C}^0([0,T])$ in an infinite dimensional vector space, it is reasonable to believe that, for ‘most’ choices of the $n$ functions $a_i$, the $(a_{i,j} a_{k,l})$ will be linearly independent. Of course, the C-NOT gate example does not satisfy this assumption, because for all $t$, $\cos^2(t) + \sin^2(t) = 1$ and also because one of the components is constant equal to 0.

\begin{proof}
\textit{1. For n=m=2.}
We first present the proof in the simplest situation $\mathcal{H} = \mathbb{C}^2 \otimes \mathbb{C}^2$, after which the argument in the general case given below will appear much clearer. Proceeding exactly as in the proof of Proposition \ref{C-NOT}, suppose that such a TPS exists, and denote $U = (u_{ij})_{1 \leq i,j \leq 4}$ the unitary that maps $\mathcal{B} = (\ket{1 1},\ket{1 2}, \ket{2 1}, \ket{2 2})$ to the basis $\mathcal{B}' = (\ket{1' 1'},\ket{1' 2'}, \ket{2' 1'}, \ket{2' 2'})$ defining this new TPS. For convenience, we also relabel $(a_{1 1}, a_{1 2}, a_{2 1}, a_{2 2})$ as $(a, b, c, d)$. In $\mathcal{B}'$, the trajectory reads:
\[U \begin{pmatrix}  a(t) \\ b(t) \\ c(t) \\  d(t) \end{pmatrix} =
\begin{pmatrix}  u_{11} a(t) + \dots + u_{14} d(t) \\  u_{21} a(t) + \dots + u_{24} d(t) \\ u_{31} a(t) + \dots + u_{34} d(t) \\  u_{41} a(t) + \dots + u_{44}d(t) \end{pmatrix}. \]
For it to be a product state, exactly as we obtained \eqref{product_state}, we must have in particular:
\begin{align}
\Big( u_{11} a(t) + u_{12} b(t) + u_{13} c(t) + u_{14} d(t) \Big) \Big( u_{41} a(t) + u_{42} b(t) + u_{43} c(t) + u_{44}d(t) \Big)  \nonumber \\
 = \Big( u_{21} a(t) + u_{22} b(t) + u_{23} c(t) + u_{24} d(t) \Big) \Big(u_{31} a(t) + u_{32} b(t) + u_{33} c(t) + u_{34} d(t) \Big) \label{cross_product_1} \end{align}
By linear independence of the $(a_i a_j)_{1\leq i,j \leq 4}$ in $\mathcal{C}^0([0,T])$, considering the coefficients in front of either $a^2$, $b^2$, $c^2$ or $d^2$, we first deduce that for all $i$,
\begin{equation} u_{1i} u_{4i} = u_{2i} u_{3i}. \label{coef} \end{equation}
Now, we distinguish between two possible cases.

\begin{itemize}
\item \textit{First case}. Suppose that there exists at least an index $i$ such that the quantity \eqref{coef} is non-zero. Up to relabelling, we can suppose without loss of generality that $u_{11} u_{41} = u_{21} u_{31} \equiv C \neq 0$. Dividing \eqref{cross_product_1} by $C$ and denoting $v_{ij} = \frac{u_{ij}}{u_{i1}}$ yields:
\begin{align*}
\Big(  a(t) + v_{12} b(t) + v_{13} c(t) + v_{14} d(t) \Big) \Big( a(t) + v_{42} b(t) + v_{43} c(t) + v_{44}d(t) \Big) \\
 = \Big( a(t) + v_{22} b(t) + v_{33} c(t) + v_{24} d(t) \Big) \Big(a(t) + v_{32} b(t) + v_{33} c(t) + v_{34} d(t) \Big) 
\end{align*}
Using linear independence again, with respect to the coefficients in front of $b^2$, $ab$, $c^2$, $ac$, $d^2$ and $ad$, we respectively get for all $i \in  \llbracket 2, 4\rrbracket$:
\[ \left\{ \begin{array}{ll}
v_{1i} v_{4i}  = v_{2i} v_{3i}  \\
v_{1i} + v_{4i}  = v_{2i} + v_{3i}.
\end{array} \right.  \]
We now multiply the second of these two equations by $v_{2i}$ and use the first one to obtain:
\begin{align*} 
 &v_{2i}  \left (v_{1i} + v_{4i} \right) =  v_{2i}^2 + v_{1i} v_{4i}  \\
\Rightarrow \quad & v_{2i}  \left (v_{1i} - v_{2i} \right) = v_{4i} \left (v_{1i} - v_{2i} \right)    \\
\Rightarrow \quad &\left (v_{1i} - v_{2i} \right) \left (v_{2i} - v_{4i} \right) = 0 \\
\Rightarrow \quad & \left\{ \begin{array}{ll} v_{1i}  = v_{2i}  \\ v_{4i}  = v_{3i} \end{array} \right. \quad \text{or} \quad \left\{ \begin{array}{ll} v_{1i}  = v_{3i}  \\ v_{4i}  = v_{2i} \end{array} \right. \quad \text{\small{(using the second equation again.)}}
\end{align*}

At first sight, it seems that having coefficients 1 paired with 2 and 3 with 4, or 1 with 3 and 2 with 4, may depend on $i$. To see that this can't happen, suppose that there exist $i_0 \neq j_0$ such that:
\[ \left\{ \begin{array}{ll} v_{1i_0}  = v_{2i_0}  \\ v_{4i_0}  = v_{3i_0} \end{array} \right. \quad \text{but} \quad \left\{ \begin{array}{ll} v_{1j_0}  = v_{3j_0}  \\ v_{4j_0}  = v_{2j_0}. \end{array} \right.  \]
By appropriately using linear independence once more, one finds that:
\[ v_{1i_0}v_{4j_0} + v_{1j_0}v_{4i_0} = v_{2i_0}v_{3j_0} + v_{2j_0}v_{3i_0}, \]
which rewrites:
\begin{align*}
&v_{1i_0}v_{4j_0} + v_{1j_0}v_{4i_0} = v_{1i_0}v_{1j_0} + v_{4j_0}v_{4i_0} \\
\Rightarrow \quad & \left(v_{1i_0} - v_{4i_0} \right)  \left(v_{4j_0} - v_{1j_0} \right) = 0 \\
\Rightarrow \quad & v_{1i_0}  = v_{2i_0} = v_{3i_0}  = v_{4i_0} \quad \text{or} \quad v_{1j_0}  = v_{2j_0} = v_{3j_0}  = v_{4j_0} \\
\Rightarrow \quad & \left[ \left\{ \begin{array}{ll} v_{1i_0}  = v_{3i_0}  \\ v_{4i_0}  = v_{2i_0} \end{array} \right. \quad \text{and} \quad \left\{ \begin{array}{ll} v_{1j_0}  = v_{3j_0}  \\ v_{4j_0}  = v_{2j_0} \end{array} \right. \right] \quad \text{or} \quad \left[ \left\{ \begin{array}{ll} v_{1i_0}  = v_{2i_0}  \\ v_{4i_0}  = v_{3i_0} \end{array} \right. \quad \text{and} \quad \left\{ \begin{array}{ll} v_{1j_0}  = v_{2j_0}  \\ v_{4j_0}  = v_{3j_0} \end{array} \right. \right],
\end{align*}
meaning that we can indeed suppose that we are in the same case for all indices, \textit{i.e.} we have (for instance, up to relabelling) for all $i$,
\[ \left\{ \begin{array}{ll} v_{1i}  = v_{2i}  \\ v_{4i}  = v_{3i}. \end{array} \right.\]
Recalling the definition of the $v_{ij} $, this implies the existence of constants $\alpha, \beta \in \mathbb{C}$ such that
\[  \begin{pmatrix} u_{11} \\ u_{12} \\ u_{13} \\  u_{14} \end{pmatrix} = \alpha \begin{pmatrix} u_{21} \\ u_{22} \\ u_{23} \\  u_{24} \end{pmatrix} \quad \text{and} \quad \begin{pmatrix} u_{31} \\ u_{32} \\ u_{33} \\  u_{34} \end{pmatrix} = \beta \begin{pmatrix} u_{41} \\ u_{42} \\ u_{43} \\  u_{44} \end{pmatrix}. \]
This, however, is impossible, because $U$ must be invertible.

\item \textit{Second case}. If, alternatively, the expression \eqref{coef} vanishes for all $i$, then the situation is actually simpler. Indeed, this means that the coefficients in front of $a^2$, $b^2$, $c^2$ and $d^2$ vanish but, looking at \eqref{cross_product_1}, this forces each function $a$, $b$, $c$ and $d$ to appear with non-zero coefficient at most once on each side of the equality (\textit{i.e.} to be present in one bracket only, on each side), otherwise a term $a^2$, $b^2$, $c^2$ or $d^2$ would appear. But, on the other hand, each function $a$, $b$, $c$ and $d$ has to appear at least in one bracket, on each side, otherwise one column of $U$ would be null. This induces, on each side of the equality, a partition of $\llbracket 1, 4\rrbracket$ into two disjoint sets of indices (functions appearing in one bracket or in the other). Furthermore, this partition is actually the same on each side. Otherwise, developing the brackets on one side would yield terms $a_i a_j$ that cannot appear on the other side, which is prevented by linear independence. Said differently, we have for all $i$, [$u_{1i} = 0 \Leftrightarrow u_{2i}=0$ and $u_{3i} = 0 \Leftrightarrow u_{4i}=0$] or [$u_{1i} = 0 \Leftrightarrow u_{3i}=0$ and $u_{2i} = 0 \Leftrightarrow u_{4i}=0$], and we can assume up to relabelling the first situation (1 paired with 2 and 3 with 4).

Now, pick a non-zero coefficient among $\{ u_{11}, u_{12}, u_{13}, u_{14} \}$ (which must exist by invertibility of $U$), say for instance $u_{11} \neq 0$ without loss of generality, and therefore $u_{21} \neq 0$ as well. By linear independence, for all $i$ we have: $u_{11} u_{3i} = u_{21} u_{4i}$, which implies that the $3^{\text{rd}}$ and $4^{\text{th}}$ lines in $U$ are proportional, again in contradiction with $U$ being invertible.
\end{itemize}

\textit{2. Arbitrary n and m.}
The above proof can be directly translated to the general case, although the notations become significantly more difficult to handle. Denote $U = (u_{(i,j) , (k,l)})_{\substack{1 \leq i,k \leq n \\ 1 \leq j,l \leq m}}$ the unitary that maps $\mathcal{B}$ to $\mathcal{B}' = (\ket{1' 1'}, \dots, \ket{n' m'})$, in which the trajectory reads:

\[U \begin{pmatrix}  a_{1,1}(t) \\ a_{1,2}(t) \\ \vdots \\  a_{n,m}(t) \end{pmatrix} =
\begin{pmatrix}  u_{(1,1),(1,1)} a_{1,1}(t) + \dots + u_{(1,1),(n,m)} a_{n,m}(t) \\  u_{(1,2),(1,1)} a_{1,1}(t) + \dots + u_{(1,2),(n,m)} a_{n,m}(t) \\ \vdots \\  u_{(n,m),(1,1)} a_{1,1}(t) + \dots + u_{(n,m),(n,m)} a_{n,m}(t) \end{pmatrix} \equiv  \begin{pmatrix} \Psi'_{1,1}(t) \\ \Psi'_{1,2}(t) \\ \vdots \\  \Psi'_{n,m}(t) \end{pmatrix}. \]
To get a product state, we must have $\Psi'_{i,j}(t) \Psi'_{k,l}(t) = \Psi'_{i,l}(t) \Psi'_{k,j}(t)$ for all $i,k \in \llbracket 1, n\rrbracket$ and $j,l \in \llbracket 1, m\rrbracket$, that is:
\begin{align}
\Big( u_{(i,j),(1,1)} a_{1,1}(t) + \dots + u_{(i,j),(n,m)} a_{n,m}(t) \Big) \Big(u_{(k,l),(1,1)} a_{1,1}(t) + \dots + u_{(k,l),(n,m)} a_{n,m}(t) \Big) \nonumber \\
 = \Big( u_{(i,l),(1,1)} a_{1,1}(t) + \dots + u_{(i,l),(n,m)} a_{n,m}(t) \Big) \Big(u_{(k,j),(1,1)} a_{1,1}(t) + \dots + u_{(k,j),(n,m)} a_{n,m}(t) \Big) \label{cross_product_2} \end{align}
We are now in the same situation as in \eqref{cross_product_1}. Instead of 4, there are now $nm$ functions, but the algebraic relations are similar. By distinguishing again between the two cases (whether at least one function $a_{i,j}^2$ has a non-zero contribution or not) and using linear independence to constrain the coefficients, one reaches the conclusion that some lines in $U$ are proportional, which is impossible.

\end{proof}

\section{The Hamiltonian perspective} \label{hamiltonian}

In the construction of Sec. \ref{existence}, in the new TPS defined by the basis $\mathcal{B}'$, the C-NOT gate evolution generates no entanglement. One therefore expects that the Hamiltonian $\hat{H}_\text{C-NOT}$ takes in this basis a simple separable form of the kind $\hat{H}_1 \otimes \bbone_2 + \bbone_1 \otimes \hat{H}_2$. This can easily be checked, because in the standard TPS defined by $\mathcal{B}$, the Hamiltonian of the gate can be chosen as: 
\[ \hat{H}_\text{C-NOT} = 
\begin{pmatrix} 
 0 & 0 & 0 & 0  \\ 
 0 & 0 & 0 & 0 \\ 
 0 & 0 & 0 & i \\
 0 & 0 & -i & 0 \end{pmatrix} \]
 (check that this indeed yields $\exp(i \hat{H}_\text{C-NOT} t) \ket{\Psi(0)} = \ket{\Psi(t)}$). Expressed in $\mathcal{B}'$, it now writes:
\[ U \hat{H}_\text{C-NOT} U^\dagger =
\begin{pmatrix} 
 0 & 1 & 1 & 0  \\ 
 1 & 0 & 0 & 1 \\ 
 1 & 0 & 0 & 1 \\
 0 & 1 & 1 & 0 \end{pmatrix} =  \begin{pmatrix}  0 & 1 \\ 1 &  0 \end{pmatrix} \otimes \bbone_2 +  \bbone_1 \otimes \begin{pmatrix}  0 & 1 \\ 1 & 0 \end{pmatrix}, \]
as expected.

The whole study presented in this work could alternatively be viewed from the Hamiltonian perspective, rather than from the vector state perspective which we have adopted. The mathematical question to solve would be: given a Hamiltonian, can we find a TPS in which the latter is of the separable form $\hat{H}_1 \otimes \bbone_2 + \bbone_1 \otimes \hat{H}_2$?

This Hamiltonian approach was notably studied by Tegmark in \cite[Sec. 3]{tegmark2015consciousness}. A Hamiltonian that can be disentangled is characterized by the fact that $\Pi_3 \hat{H} = 0$, following to the notations of the reference (equation (24))\footnote{However, the proof of the main result given there, the ‘Diagonality theorem’ according to which any Hamiltonian is maximally separable in its eigenbasis, seems insufficient to conclude as it stands. Indeed, the author looks for changes of TPS that minimize the distance from $\hat{H}$ to the subspace of perfectly separable operators. To do so, he requires the differential of this distance to vanish (equation (35) in the reference), but this is only a necessary condition for being minimal, since there could be local minima. Our C-NOT example works fine, though, as $\hat{H}_\text{C-NOT}$ is indeed separable in its eigenbasis, because in the latter it reads (up to a permutation): 
\[ \begin{pmatrix} 
1 & 0 & 0 & 0  \\ 
 0 & 0 & 0 & 0 \\ 
 0 & 0 & 0 & 0 \\
 0 & 0 & 0 & -1 \end{pmatrix}
 = \begin{pmatrix}  1/2 & 0 \\ 0 & -1/2 \end{pmatrix} \otimes \bbone_2 +  \bbone_1 \otimes \begin{pmatrix}  1/2 & 0 \\ 0 & -1/2 \end{pmatrix} \]}.

We can answer this question in more generality in the case of a TPS composed of $n$ finite dimensional factors. A Hamiltonian that admits a disentangling TPS in which it takes the separable form $\hat{H}_1 \otimes \dots \otimes \bbone_n + \dots + \bbone_1 \otimes \dots \otimes \hat{H}_n$ is characterized by the following proposition:

\begin{proposition} \label{characterization}
A finite-dimensional Hermitian operator $\hat{H}$ admits a disentangling TPS $\mathcal{H} = \bigotimes_{i=1}^n \mathcal{H}_i$ if and only if there exist sets of real numbers $(E_i)_{1 \leq i \leq n}$ with $\lvert E_i \rvert = \dim(\mathcal{H}_i)$ such that the spectrum of $\hat{H}$ is of the form:
\[ \mathrm{spec}(\hat{H}) = ( \lambda_1 + \dots + \lambda_n )_{ \lambda_i \in E_i}, \]
respecting the multiplicities. In this case, the eigenvectors of $\hat{H}$ are product states for the disentangling TPS.
\end{proposition}

\begin{proof}
For the direct implication, let's write $\hat{H} = \hat{H}_1 \otimes \dots \otimes \bbone_n + \dots + \bbone_1 \otimes \dots \otimes \hat{H}_n$ in the disentangling TPS, and observe that the vectors $( \ket{e_1^{\lambda_1}} \otimes \dots \otimes \ket{e_n^{\lambda_n}} )_{\lambda_i \in \mathrm{spec}(\hat{H}_i)}$ (where $\ket{e_i^{\lambda_i}} \in \mathcal{H}_i$ denotes the eigenvector of $\hat{H}_i$ associated with the eigenvalue $\lambda_i$) are eigenvectors of $\hat{H}$ associated with the eigenvalues $( \lambda_1 + \dots + \lambda_n )_{ \lambda_i \in E_i}$. There are $ \prod_{i=1}^n \dim(\mathcal{H}_i) = \dim(\mathcal{H})$ such vectors, so they form a complete eigenbasis.

Conversely, if $\hat{H}$ has a spectrum of the form $( \lambda_1 + \dots + \lambda_n )_{ \lambda_i \in E_i}$, then for all $(\lambda_1, \dots,  \lambda_n) \in E_1 \times \dots \times E_n$, we label by $\ket{e_1^{\lambda_1} \dots e_n^{\lambda_n}} \in \mathcal{H}_i$ the eigenvector of $\hat{H}$ associated with the eigenvalue $ \lambda_1 + \dots + \lambda_n$. This mere notation defines in fact a TPS $\mathcal{H} = \bigotimes_{i=1}^n \mathcal{H}_i$. We then call $\hat{H}_i$ the operator on $\mathcal{H}_i$ whose eigenvectors are $(\ket{e_i^{\lambda_i}})_i$ associated respectively with the eigenvalues $(\lambda_i)_i$. Since $\hat{H}$ and $\hat{H}_1 \otimes \dots \otimes \bbone_n + \dots + \bbone_1 \otimes \dots \otimes \hat{H}_n$ coincide on the eigenbasis of $\hat{H}$, they are equal as operators.
\end{proof}

\section{Conclusion}

As an attempt to push further some questions arising in the growing field of quantum mereology, we have introduced the notion of disentangling tensor product structure (TPS), defined as a factorization of a Hilbert space in which a given time-evolving quantum state is a product state at all times. In the case of a C-NOT gate evolution between two qbits, we were able to explicitly construct a disentangling TPS. However, we have shown that such a construction is not possible in general under a very weak assumption on the time-evolving quantum state, leading to the idea that ‘most’ quantum systems cannot be disentangled. In the language of \cite{franzmann2024or}, this result may be seen as the analog of the fact that normal coordinates can only be found for an infinitesimally small region in spacetime.

The next natural step would be to search for approximate disentanglement. This amounts to determine the TPS $\mathcal{T}$ for which a given  $\ket{\Psi(t)}$ deviates as little as possible from the set of the product states $P_\mathcal{T}$, \textit{i.e.} the TPS that minimizes $\underset{t \in [0,T]}{\sup} d( \ket{\Psi(t)} , P_\mathcal{T})$ for some suitable distance $d$. Some ideas are perhaps to be found in \cite{wiesniak2020distance}, but this quest seems fairly unreachable, in particular if the factorization of $\mathcal{H}$ includes not only two but $n$ tensor factors, due to the extremely complex nature of the set $P_\mathcal{T}$. Determining whether a vector belongs to $P_\mathcal{T}$ is known to be a NP-hard problem, in fact most tensor problems are \cite{hillar2013most}.

Perhaps the Hamiltonian approach explored in Sec. \ref{hamiltonian} is a more promising route. Instead of looking for a TPS that turns $\ket{\Psi(t)}$ into an approximate product state, we could look for a TPS that minimizes the interaction part of the Hamiltonian. Proposition  \ref{characterization} is a first step in this direction, and it might also be worth to attempt to complete the proof of Tegmark's ‘Diagonality theorem’ \cite{tegmark2015consciousness} or take inspiration from it.

\newpage
\bibliographystyle{siam}
\bibliography{Biblio_disentangling}
\end{document}